\documentclass{cernrep} 
\usepackage{texnames}
\usepackage[T1]{fontenc}
\usepackage{bm,color}
\begin{document}
\title{Fission Dynamics of Compound Nuclei}
 
\author{Yoritaka Iwata and Sophia Heinz}

\institute{GSI Helmholtzzentrum f\"ur Schwerionenforschung, Darmstadt, Germany}

\maketitle 

\begin{abstract}
Collisions between $^{248}$Cm and $^{48}$Ca are systematically investigated by time-dependent density functional calculations with evaporation prescription.
Depending on the incident energy and impact parameter, fusion, deep-inelastic and quasi-fission events are expected to appear.
In this paper, possible fission dynamics of compound nuclei is presented.
\end{abstract}
 
\section{Introduction}
The synthesis of superheavy chemical elements~\cite{07hofmann,06oganessian} in the laboratory is one of the biggest challenges in nuclear physics.
It is an attempt for clarifying the existence limits of all the chemical elements, as well as the completion attempt of the periodic table of chemical elements.
We are concerned with heavy-ion collisions
\[  ^{248}{\rm Cm}~+~^{48}{\rm Ca} \]
with different impact parameters in this paper.
Let $A$ and $Z$ be the mass number and the proton number, respectively.
The neutron number $N$ is defined by $N = A -Z$, so that $N/Z$ of $^{248}$Cm and $^{48}$Ca are 1.58 and 1.40, respectively.
If fusion appears, $^{296}{\rm Lv}$ (= $^{296}116$) with $N/Z = 1.55$ is produced.

Fast charge equilibration~\cite{10iwata} is expected to appear in low-energy heavy-ion reactions with an incident energy of a few MeV per nucleon.
It provides a very strict limitation for the synthesis of superheavy elements.
Actually, the $N/Z$ of final product is not above nor below the $N/Z$ of the projectile and the target (1.40 $\le$ N/Z $\le$ 1.58 in this case) in the case of charge equilibration, so that the proton-richness of the final product follows. 
Although the actual value of $N/Z$ depends on the two colliding ions, its value for the merged nucleus tends to be rather proton-rich for a given proton number of the merged system.
This feature is qualitatively understood by the discrepancy between the $\beta$-stability line and the $N=Z$-line for heavier cases. 
Superheavy compound nuclei are very fragile and fission is a very frequent channel which leads to disintegration of the compound nuclei even at low excitation energies.
In this paper, following the evaporation prescription shown in Ref.~\cite{12iwata}, a possible fission dynamics of compound nuclei is simulated based on the time-dependent density functional theory (TDDFT).

\section{Methods} \label{sec1}
\subsection{Treatment of the thermal property}
Self-consistent time-dependent density functional calculations are employed in this paper.
TDDFT reproduces the quantum transportation due to the collective dynamics. 
In this sense, what is calculated by the TDDFT can be regarded as products after several $10^{-21}$~s, which corresponds to a typical time-scale of low-energy heavy-ion reactions (1000~fm/c), as well as to the inclusive time interval of any collective oscillations such as giant dipole resonance, giant quadrupole resonance and so on.
Meanwhile, thermal properties such as the thermal instability are not directly taken into account in TDDFT.
Indeed, the Skyrme type interaction used in TDDFT (for example, see Ref.~\cite{greiner-maruhn}) is determined only from several densities.
It is important that the most effective cooling effect arises from the emission of particles, and therefore it is expected that the break-up or fission of fragments including rather high internal excitation energy is suppressed in the TDDFT final products.
The additional thermal effects leading to the break-ups of fragments should be introduced.

Here is a fact that simplifies the treatment of thermal effects, that is, the difference of the time-scales.
Different from the typical time scale of low-energy heavy-ion reactions, the typical time-scale of the thermal effects is estimated by the typical time interval of collision-fission (fission appearing in heavy-ion collisions): several 10$^{-19}$~s.
It is reasonable to introduce an evaporation prescription simply to the TDDFT final products.
In this context the TDDFT final fragments have the meaning of products just after the early stage of heavy-ion reactions (several $10^{-21}$~s).

\subsection{Evaporation prescription}

\begin{figure}
\begin{center}
\includegraphics[width=10cm]{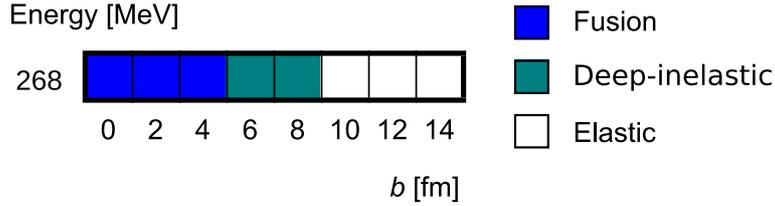}
\end{center}
\caption{\label{fig4a} (Colour online) Diagram of different reaction channels obtained in $^{48}{\rm Ca} + ^{248}{\rm Cm}$ collisions by TDDFT calculations.
The preferred reaction channels for different beam energies and impact paramters are given. The beam energies are all located above the
Coulomb barrier which is 209.0~MeV. The results show fusion, deep-inelastic and elastic events. The difference between fusion-fission 
and quasi-fission is defined in this paper by whether the fission products satisfy ``1.40 $\le$ N/Z $\le$ 1.58'' or not.}
\end{figure}

In complete fusion reactions the cross-section for the formation of a certain evaporation residue is usually given by three factors \cite{anto}:
\begin{equation}
\sigma_{ER}(E_{cm}) = \sum_J \sigma_{CP} (E_{cm},J) \times P_{CN} (E_{cm},J) 
 \times P_{SV} (E_{cm},J)
\end{equation}
where $\sigma_{CP}$, $P_{CN}$ and $ P_{SV}$ mean the capture cross-section, the probability for the compound nucleus formation, and the probability for survival of the compound nucleus against fission.
All three factors are functions of the centre-of-mass energy $E_{cm}$ and the total angular momentum $J$, where $J$ can be related with the impact parameter. 
For light systems $P_{CN}$ and $P_{SV}$ are about unity and $\sigma_{ER}\,\approx\,\sum_J\sigma_{CP}$. 
But in superheavy systems the strong Coulomb repulsion and large angular momenta lead to small values of $P_{CN}$ and $P_{SV}$ and therefore to the small cross-sections of the evaporation residues observed in the experiments. 
This means, different from light systems, it is necessary to introduce additional thermal effects for the superheavy element synthesis.
First, $\sigma_{CP}$ is sufficiently considered in the TDDFT if we restrict ourselves to a sufficiently high energy exceeding the Coulomb barrier (cf. sub-barrier effects such as tunnelling are not taken into account in the TDDFT).
Second, $P_{CN}$ is fully considered in the TDDFT, which is a kind of mass equilibration also related to charge equilibration.
Third, $P_{SV}$ whose relative time-scale is by no means equal to the former two probabilities is not satisfactorily considered in the TDDFT.
This probability is much more related to thermal effects.
Consequently, further consideration is necessary only for $P_{SV}$ as far as the energy above the Coulomb barrier is concerned.

Several factors are included in $P_{SV}$ such as probabilities for fission of the compound nucleus,
neutron-evaporation, proton-evaporation, deuteron-evaporation, alpha-particle-evaporation and so on.
Probabilities of neutron and $\alpha$-particle emissions are considered by
\[ P_{SV} :=(1- P_{n, evap}) (1- P_{\alpha, evap}). \]
For the details of evaporation prescription, see our preceding research summarized in Ref.~\cite{12iwata}.

\begin{figure}
a) ${\it b} = 4$~fm  \\
\includegraphics[width=3.7cm]{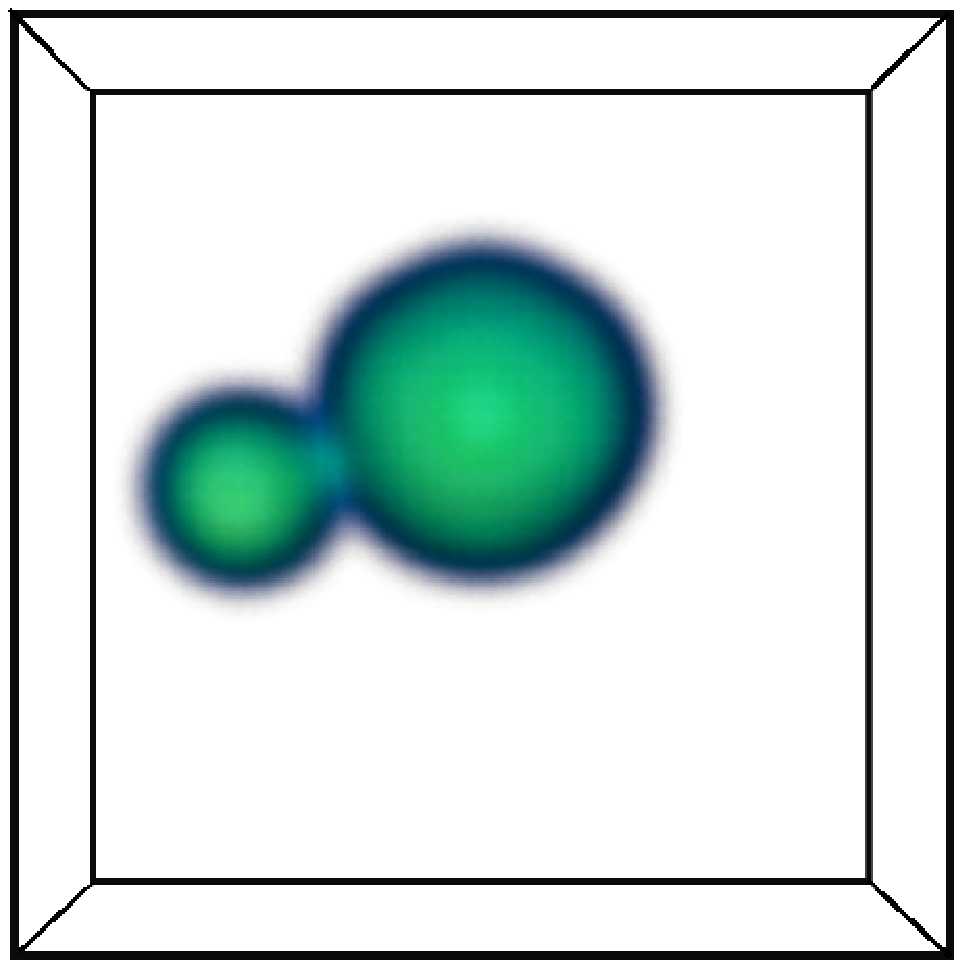}
\includegraphics[width=3.7cm]{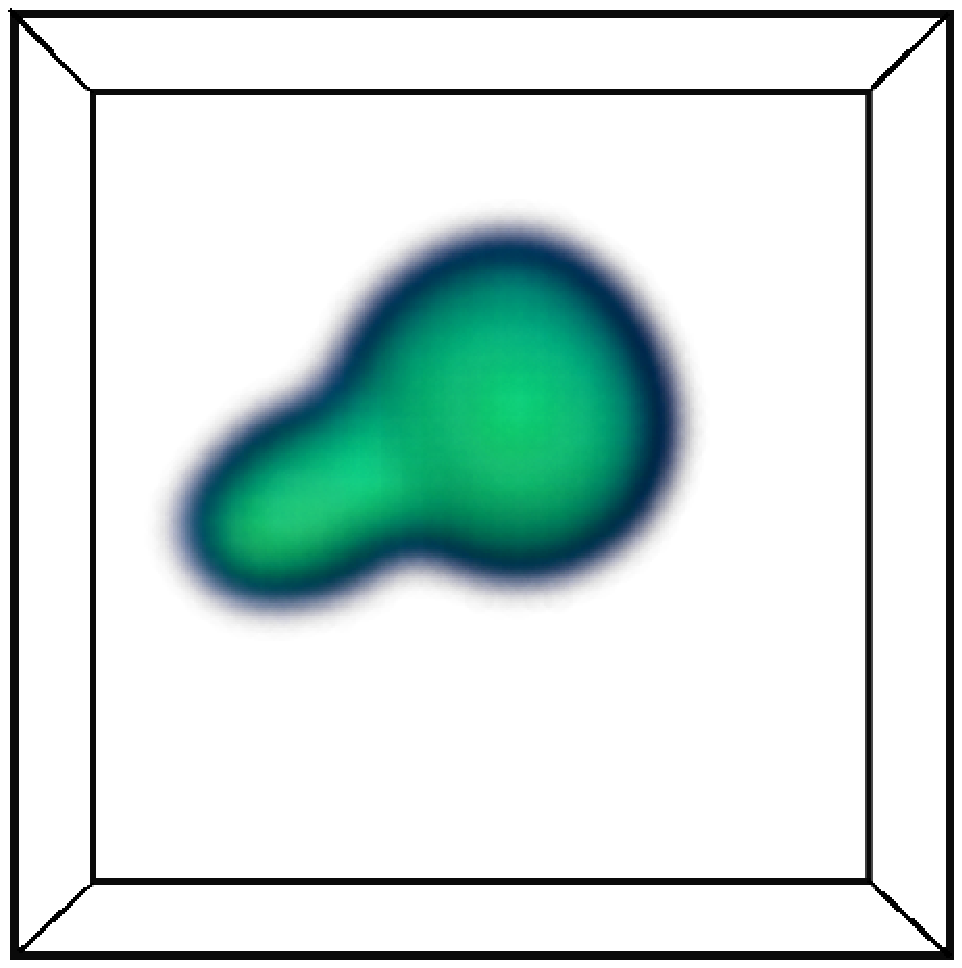}
\includegraphics[width=3.7cm]{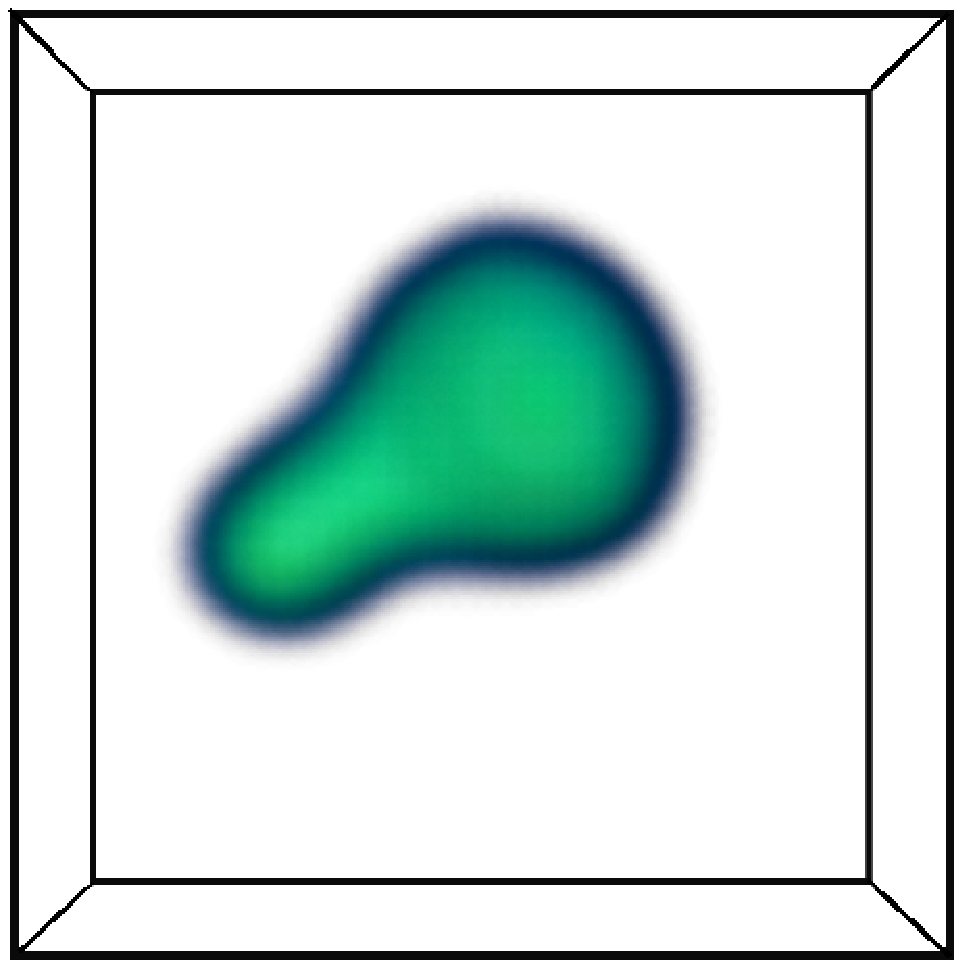}
\includegraphics[width=3.7cm]{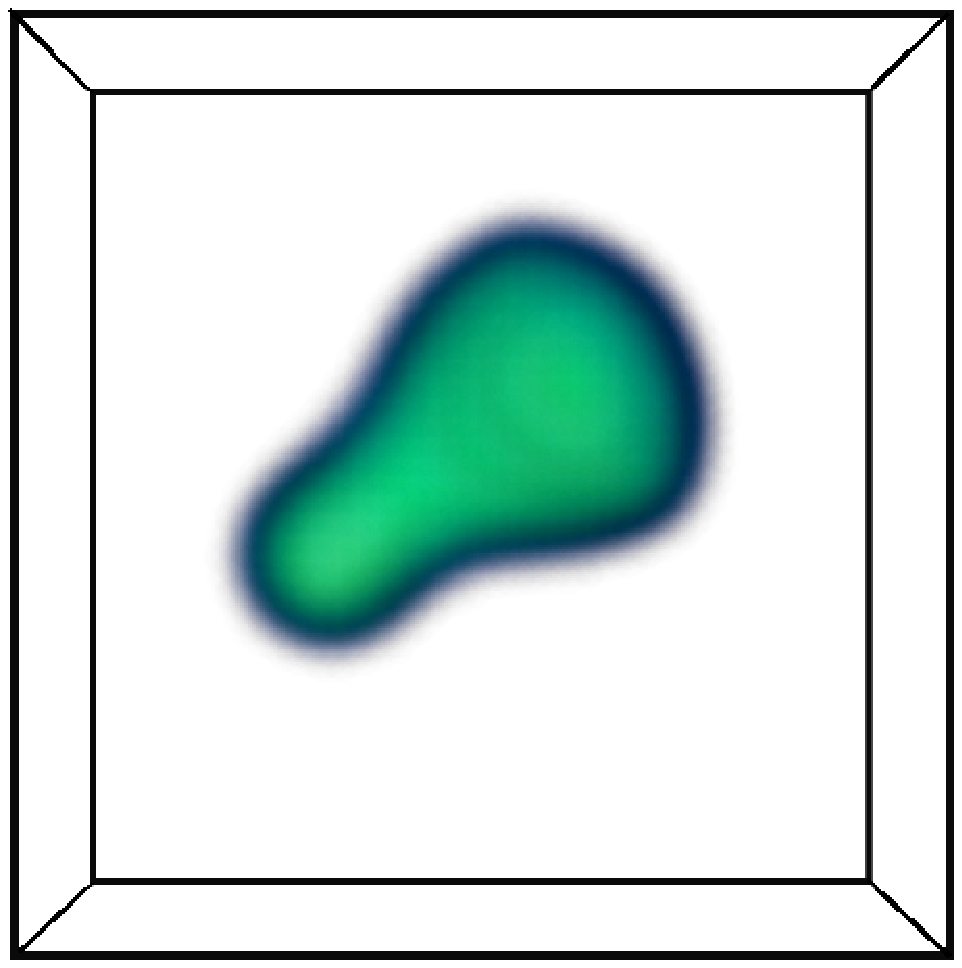}\\
b) ${\it b} = 6$~fm  \\ 
\includegraphics[width=3.7cm]{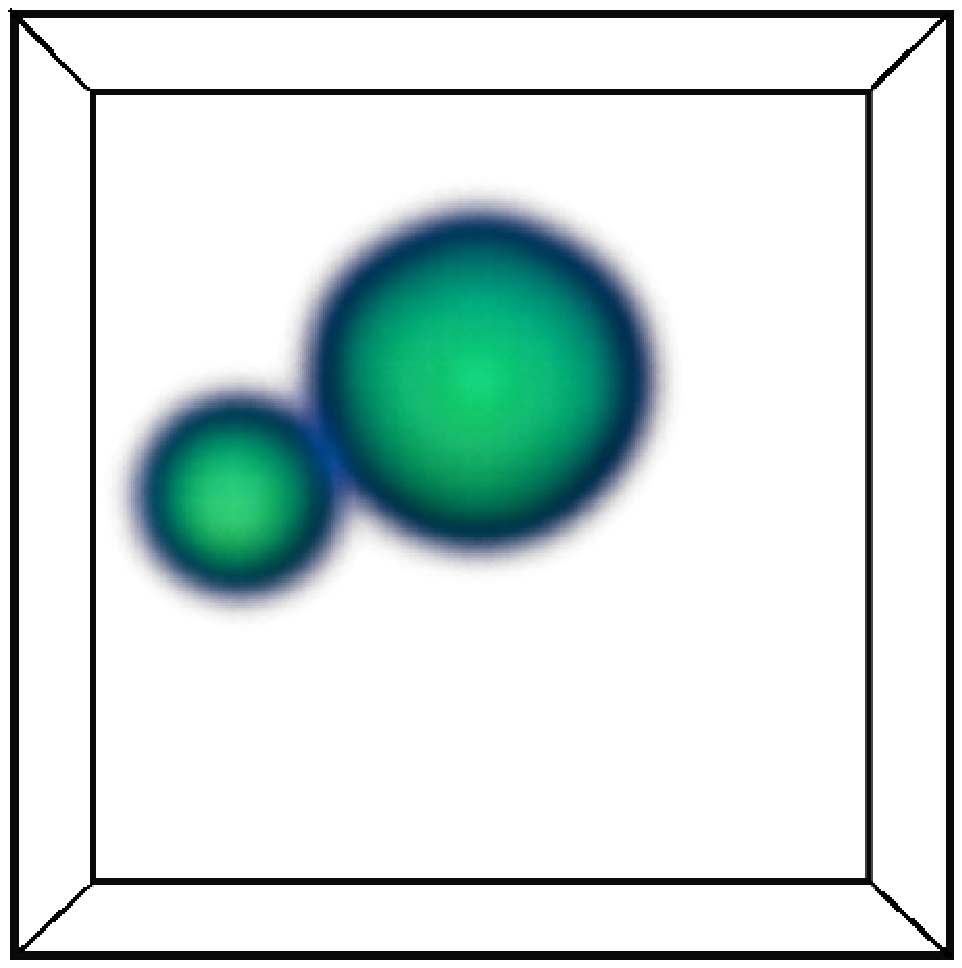}
\includegraphics[width=3.7cm]{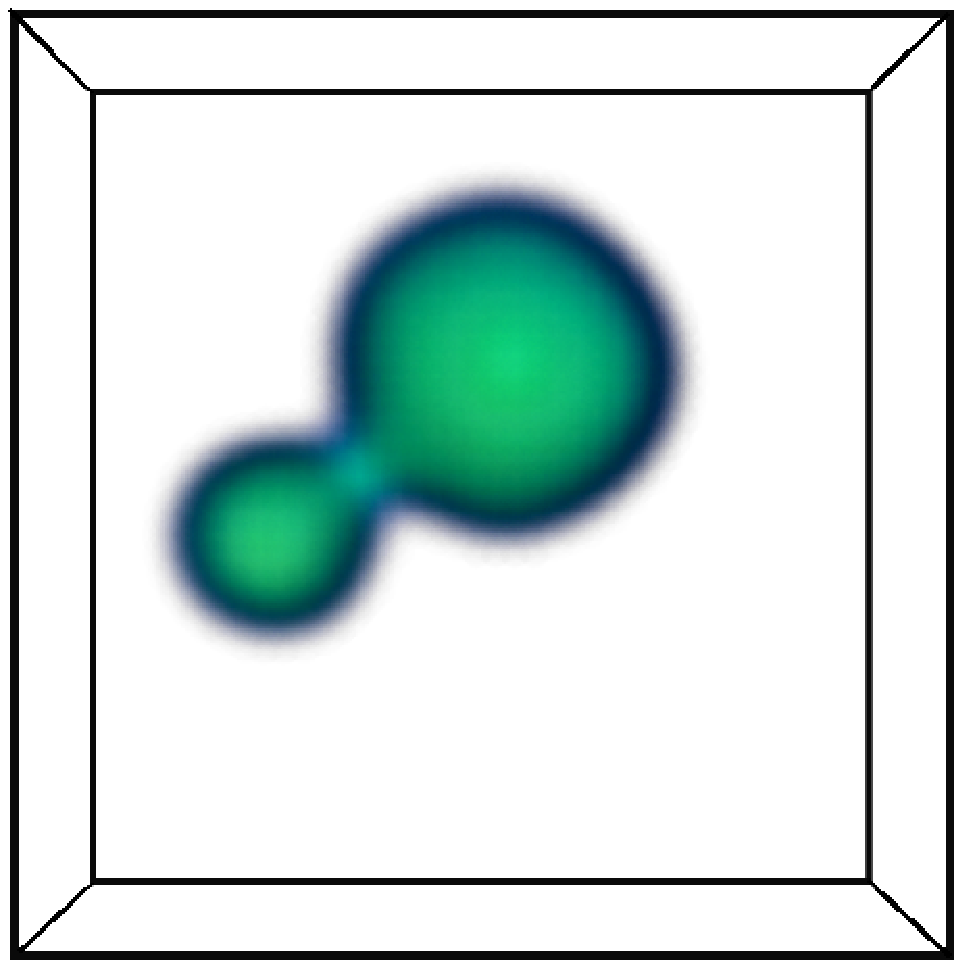}
\includegraphics[width=3.7cm]{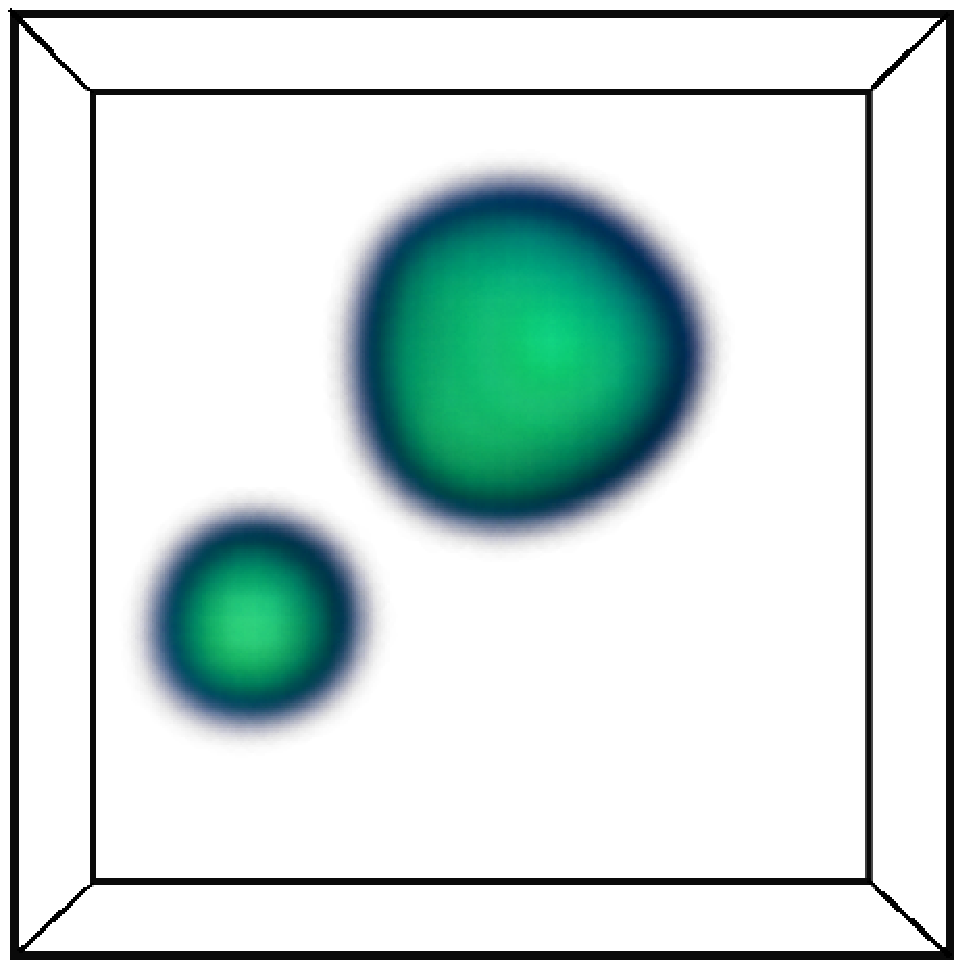}
\includegraphics[width=3.7cm]{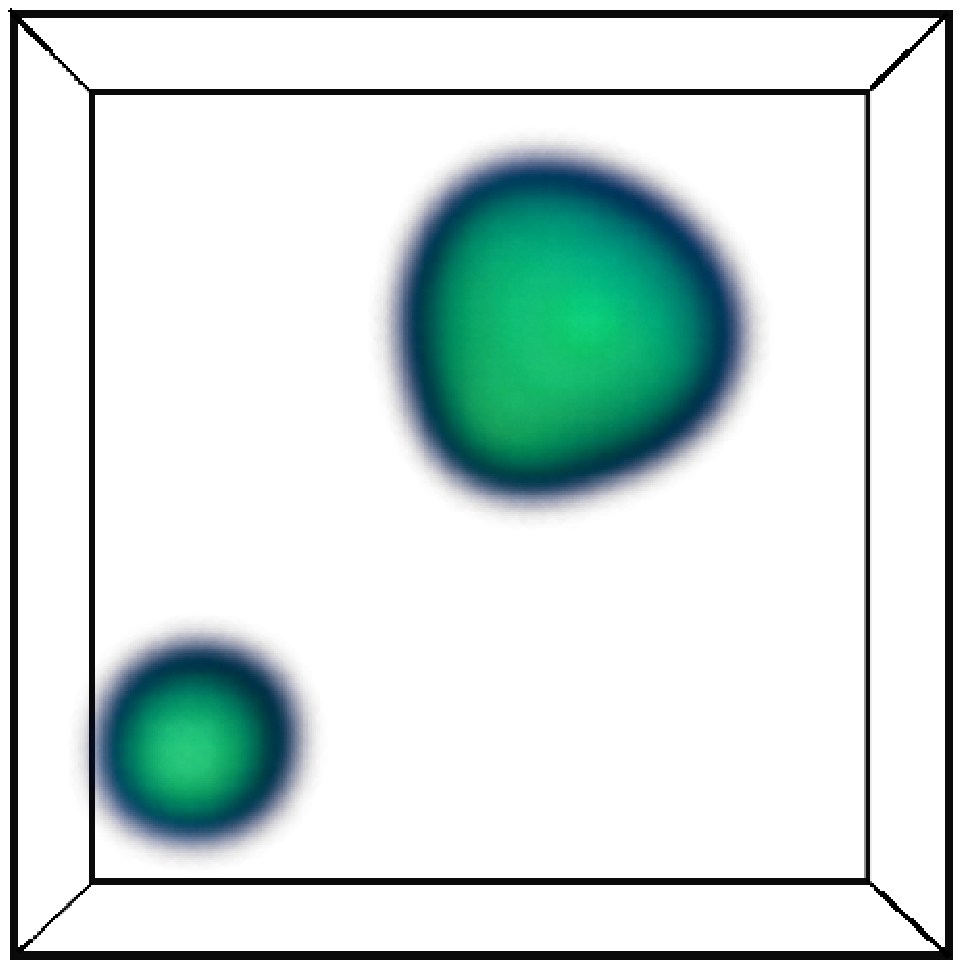}\\
c) ${\it b} = 8$~fm  \\ 
\includegraphics[width=3.7cm]{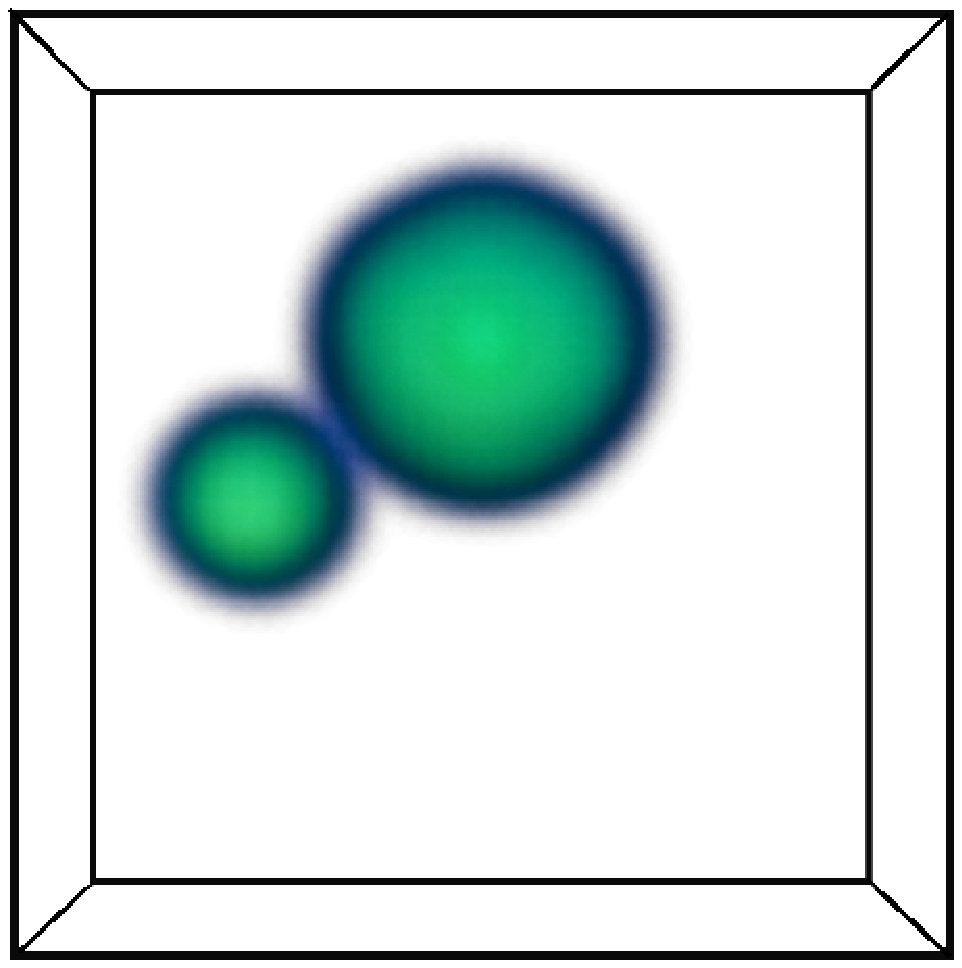}
\includegraphics[width=3.7cm]{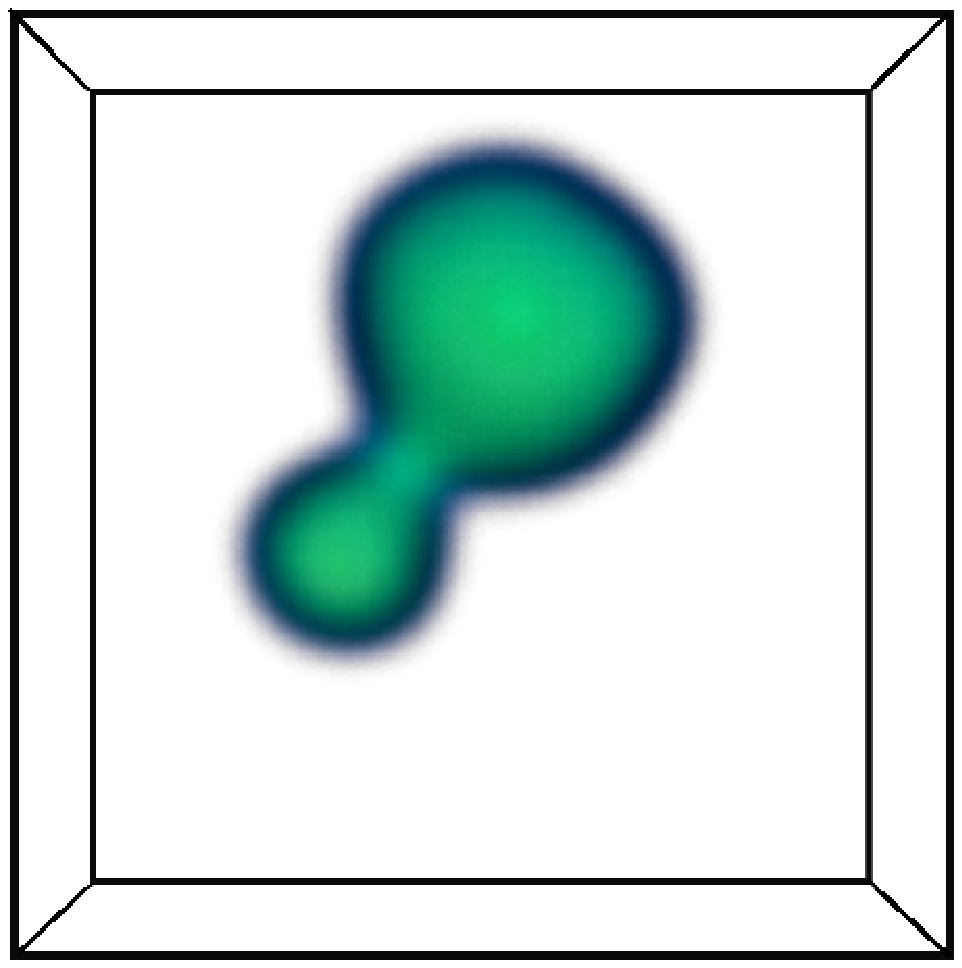}
\includegraphics[width=3.7cm]{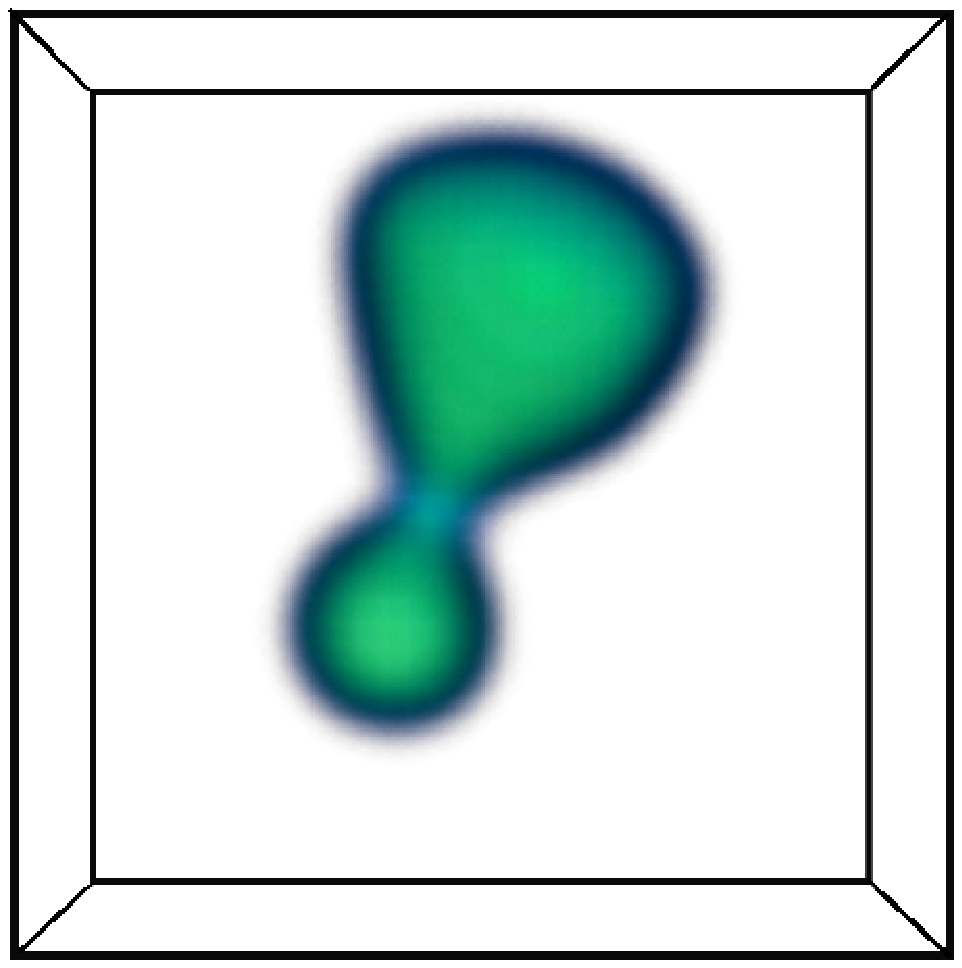}
\includegraphics[width=3.7cm]{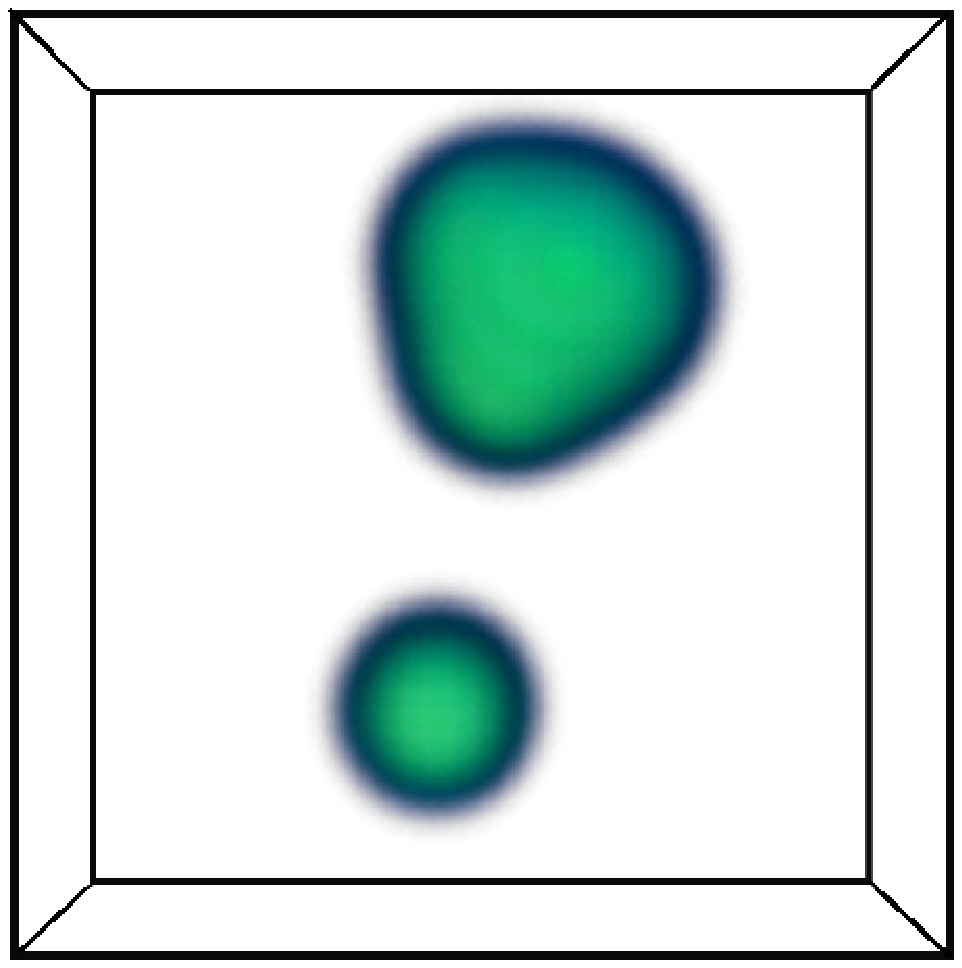}\\
d) ${\it b} = 10$~fm  \\
\includegraphics[width=3.7cm]{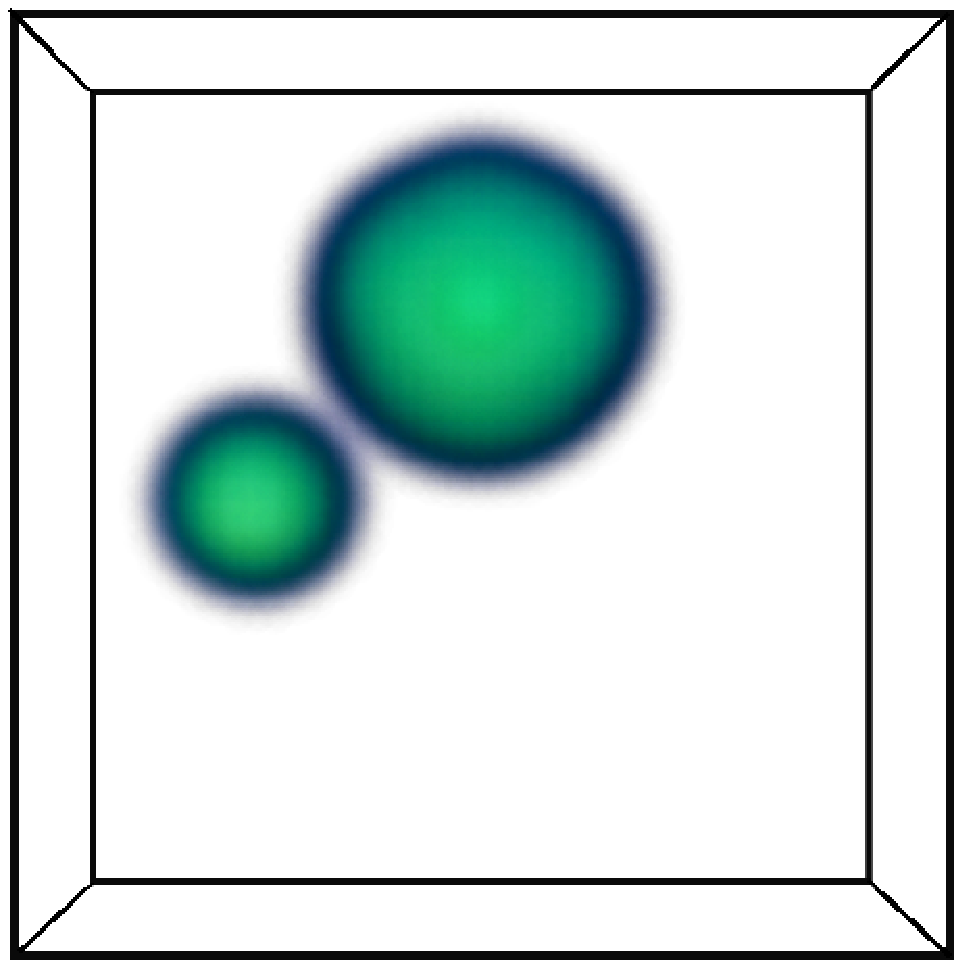}
\includegraphics[width=3.7cm]{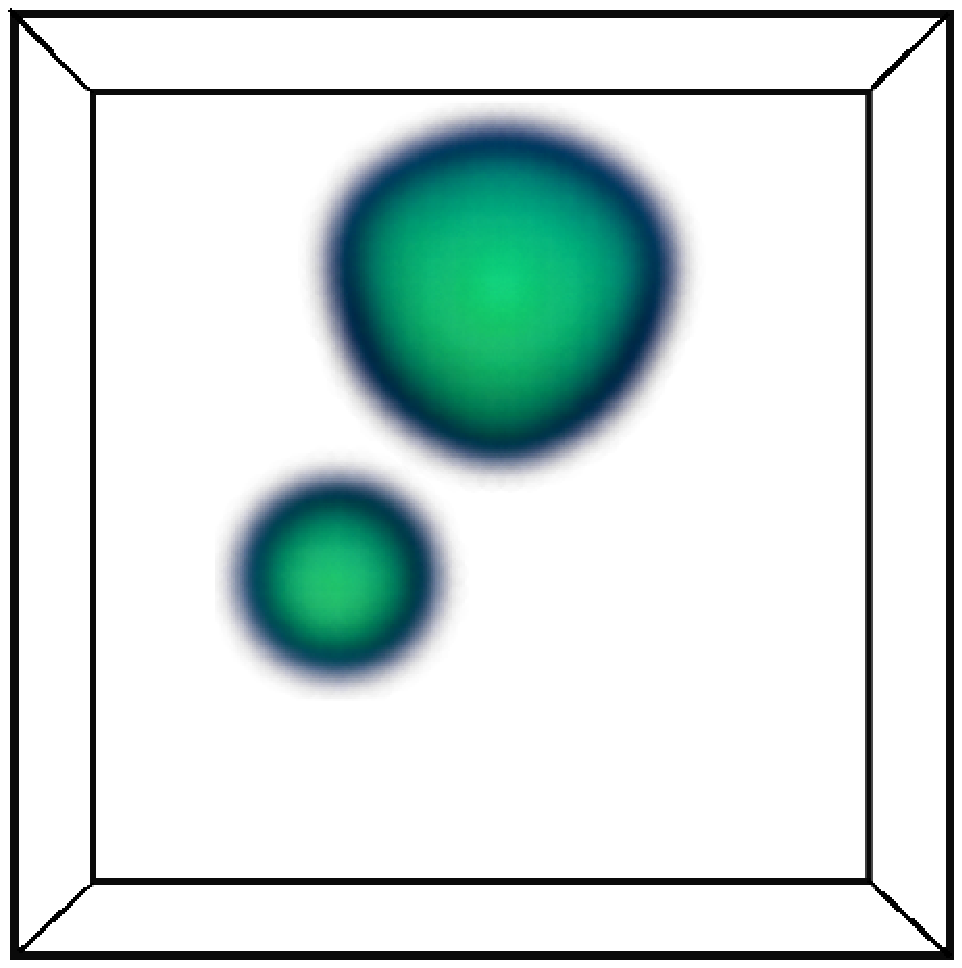}
\includegraphics[width=3.7cm]{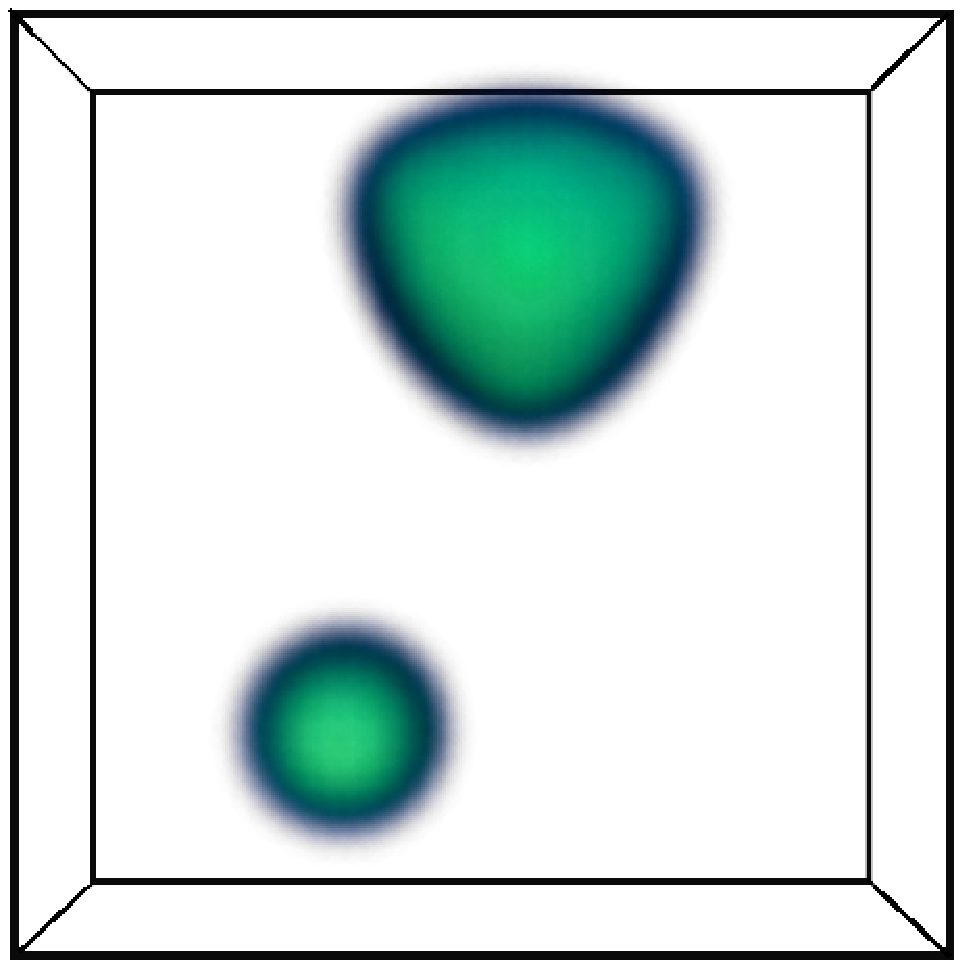}
\includegraphics[width=3.7cm]{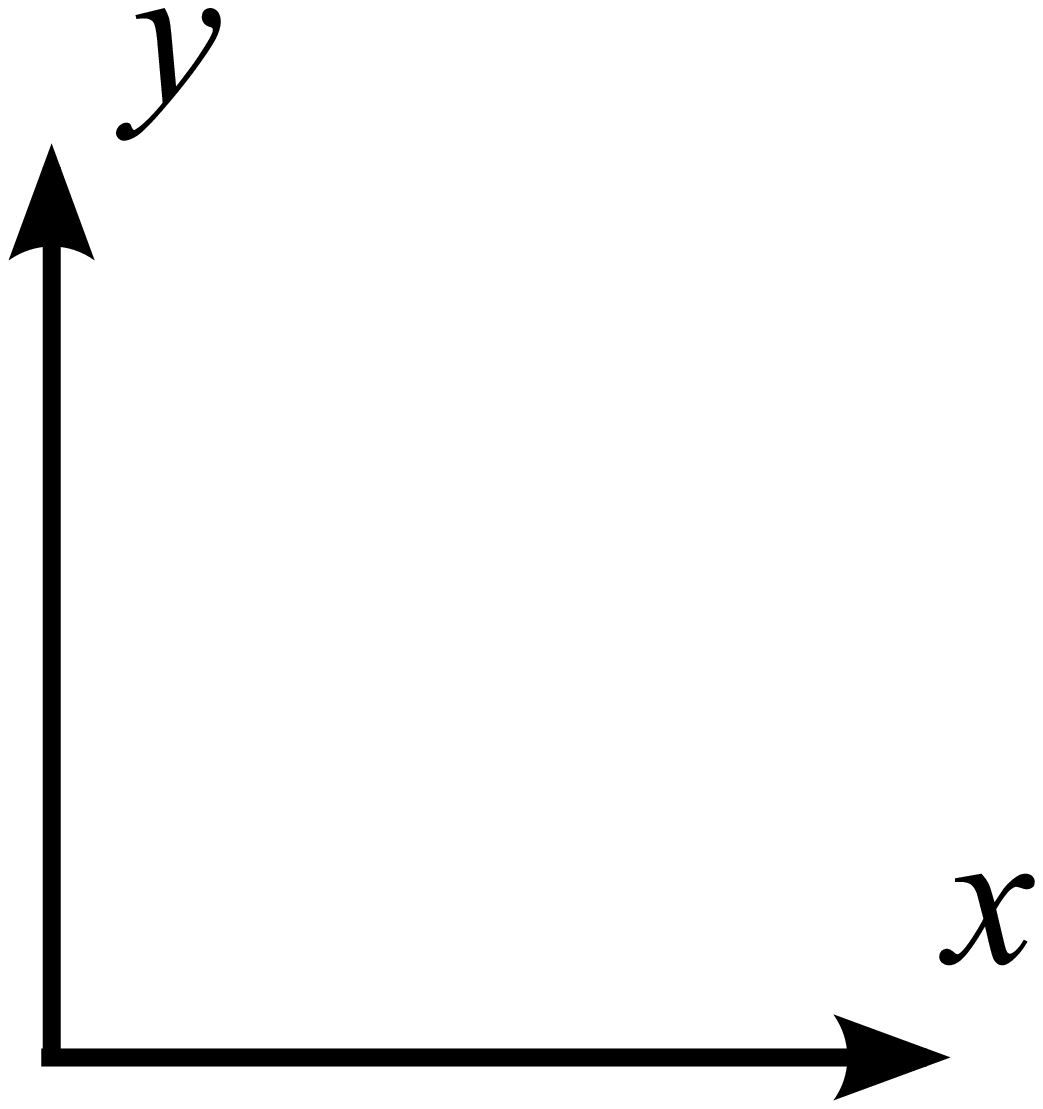}\\
\caption{\label{fig5a} (Colour online) Time evolution of $^{248}{\rm Cm}+^{48}{\rm Ca}$ with a fixed incident energy 268 MeV in the centre-of-mass frame, where a box is fixed to 48$\times$48$\times$24~fm$^3$.
Snapshots at 2.3$\times$10$^{-22}$~s, 8.7$\times$10$^{-22}$~s, 15.1$\times$10$^{-22}$~s and 21.5$\times$10$^{-22}$~s are shown, where a snapshot at 21.5$\times$10$^{-22}$~s is not shown only for ${\it b} = 10$~fm.
Fusion, deep-inelastic and elastic events appear depending on $b$.
In particular, the life-time of composite nucleus for $b=6$ and 8~fm is the order of $\times$10$^{-21}$~s.}
\end{figure}

\subsection{Fission dynamics}
For given energy and impact parameter, long-lived compound nucleus with certain excitation energies (compared to the ground state) are obtained.
Fission dynamics is obtained by additional TDDFT calculations, where no initial velocity is given.
Let us take a residual nucleus of mass number $A_R$, proton number $Z_R$ and excitation energy $E_R$.
Consider the binary fission:
\begin{equation} \label{fission}
^{A_R}Z_R \to ^{A_1}Z_1 ~+~ ^{A_2}Z_2, 
\end{equation}
where $ A_R = \sum_{i =1}^2 A_i$ and $ Z_R = \sum_{i =1}^2 Z_i$.
First, choose $A_i$ and $Z_i$ to determine the heavy-ion reaction being considered.
A configuration of the two nuclei at a distance $R_0$ is prepared as an initial state of additional calculation.
Second, choose $R_0$ such that the excitation energy agrees with that of the compound state found in the collision (Fig.~\ref{fig5a}), where the energy can be different depending on $R_0$.
Note here that the TDDFT is a theory in which the total energy is strictly conserved, so that the total energy is conserved during the presented fission process. 
Third, the initial many-body wave function, which is given as a single Slater determinant, consists of single wave functions of two different initial nuclei, where a set of single wave functions are orthogonalized before starting TDDFT calculations (cf. the Gram-Schmidt orthogonalization method).
In this way a configuration at the same excitation energy but closer to fission can be obtained; for given $A_R$, $Z_R$ and $E_R$.
The distance $R_0$ is uniquely determined for fixed $^{A_1}Z_1$ and $^{A_2}Z_2$. 

\section{TDDFT calculations for fission dynamics}  \label{sec2}
Time-dependent density functional calculations with a Skyrme interaction (SLy6~\cite{Chabanat-Bonche}) are carried out in a spatial box of $48 \times 48 \times 24$~fm$^3$ with periodic boundary condition.
The unit spatial spacing and the unit time spacing are fixed to 1.0~fm and $2/3 \times 10^{-24}$~s, respectively.

The initial positions of $^{248}$Cm and $^{48}$Ca are fixed to $(0,b,0)$ and (-15,0,0), respectively.
The initial $^{248}$Cm is almost spherical the diameter for $x$, $y$, and $z$ directions are 19~fm, 19~fm and 18~fm, respectively. 
$^{248}{\rm Cm}$ (the right hand side on $x-y$-plane) does not have initial velocity on the frame, while the initial velocity parallel to the $x$-axis is given to $^{48}{\rm Ca}$ (the left hand side on $x-y$-plane).
The systematic results of TDDFT calculations for a given incident energy (268~MeV) are summarized in Fig.~\ref{fig4a}.
These results, which include many fusion events, provide a quite optimistic view for producing superheavy elements.
However, in comparison with experiments, the corresponding fusion cross-section of those low-energy heavy-ion reactions is too high to believe.
Consequently, although these TDDFT results are still legitimate to show products just after the early stage of heavy-ion reactions, it is necessary to take into account $P_{SV}$ in order to have comparable results to experiments.
 
In case of the incident energy 268~MeV, the pure TDDFT results show the following reactions (Fig.~\ref{fig5a}):
\[ \begin{array}{ll}
 ^{248}{\rm Cm}~+~^{48}{\rm Ca} ~\to~ ^{296}{\rm Lv}    \vspace{1.5mm} \\
 ^{248}{\rm Cm}~+~^{48}{\rm Ca} ~\to~ ^{247}{\rm Cf}  +^{49}{\rm Ar}   \vspace{1.5mm} \\
 ^{248}{\rm Cm}~+~^{48}{\rm Ca} ~\to~ ^{246}{\rm Bk}  +^{50}{\rm K}   \vspace{1.5mm} \\
 ^{248}{\rm Cm}~+~^{48}{\rm Ca} ~\to~ ^{248}{\rm Cm}~+~^{48}{\rm Ca}
\end{array} \]
for $b = 4, 6, 8, 10$~fm, respectively.
If we take into account the neutron and alpha emissions, they become
\[ \begin{array}{ll}
 ^{248}{\rm Cm}~+~^{48}{\rm Ca}  ~\to~ ^{280}{\rm Ds}  + 3 \alpha + 4 n   \vspace{1.5mm} \\
 ^{248}{\rm Cm}~+~^{48}{\rm Ca}  ~\to~ ^{238}{\rm Pu}  +^{47}{\rm Ar} + 2 \alpha + 3  n   \vspace{1.5mm} \\
^{248}{\rm Cm}~+~^{48}{\rm Ca} ~\to~ ^{237}{\rm Np}  +^{48}{\rm K} + 2 \alpha + 3 n  \vspace{1.5mm} \\
 ^{248}{\rm Cm}~+~^{48}{\rm Ca} ~\to~ ^{248}{\rm Cm}~+~^{48}{\rm Ca}
\end{array} \]
for $b = 4, 6, 8, 10$~fm, respectively.

\begin{figure}
\includegraphics[width=3.7cm]{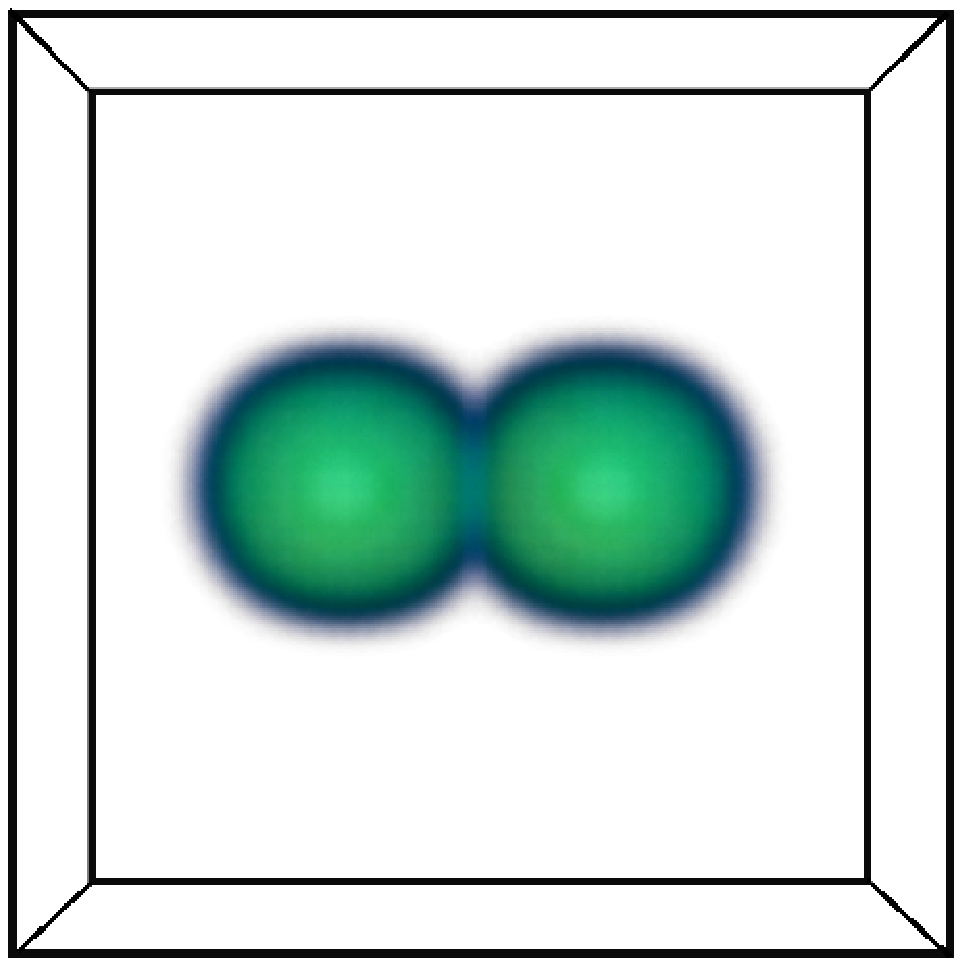}
\includegraphics[width=3.7cm]{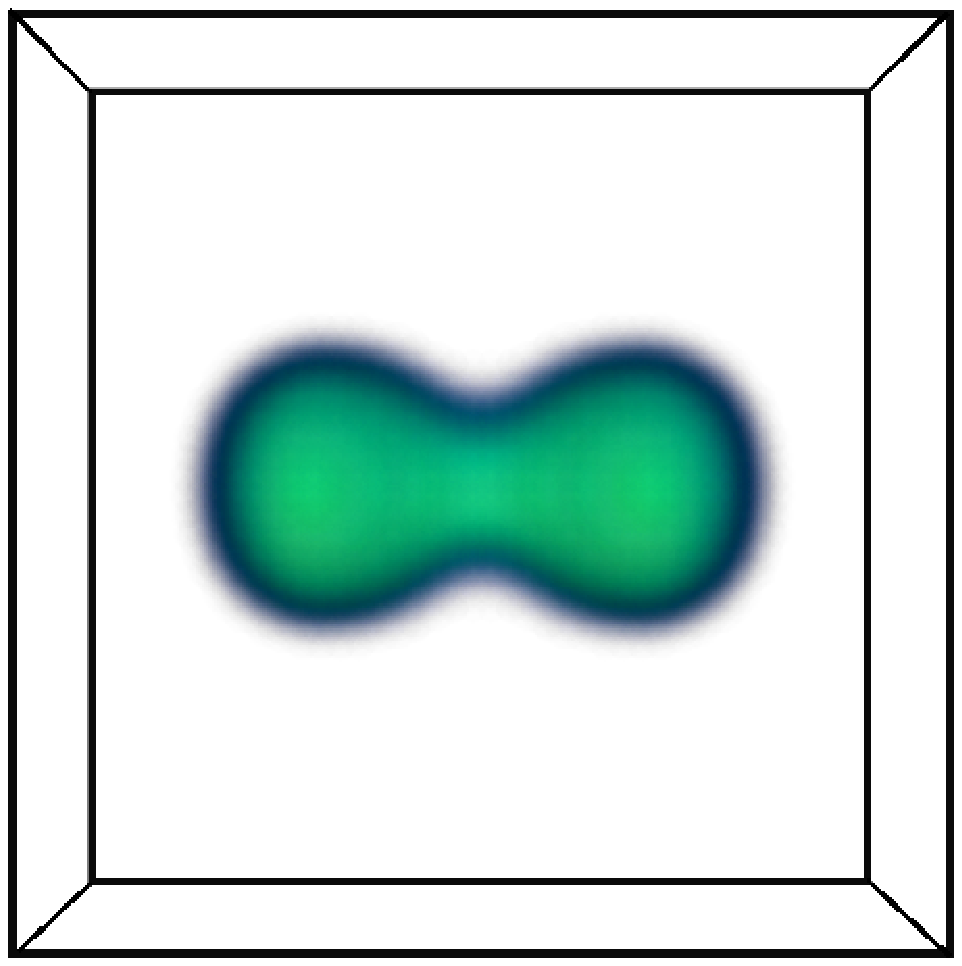}
\includegraphics[width=3.7cm]{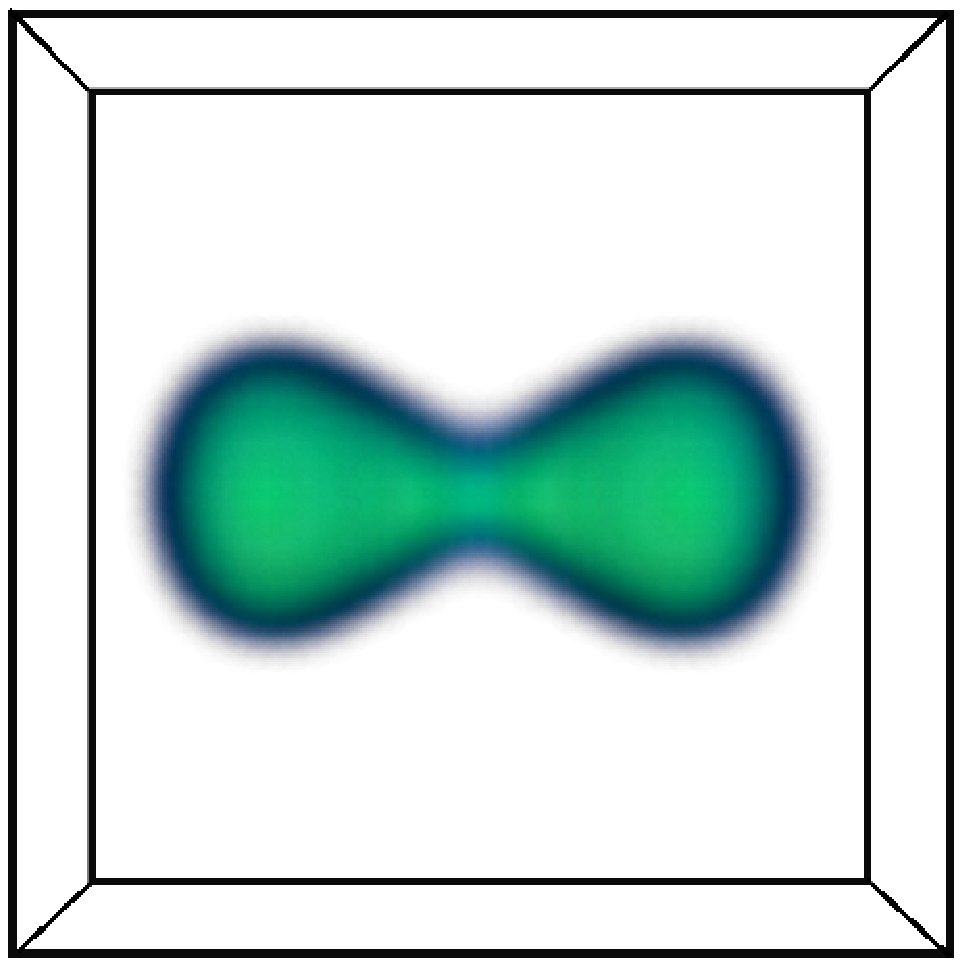}
\includegraphics[width=3.7cm]{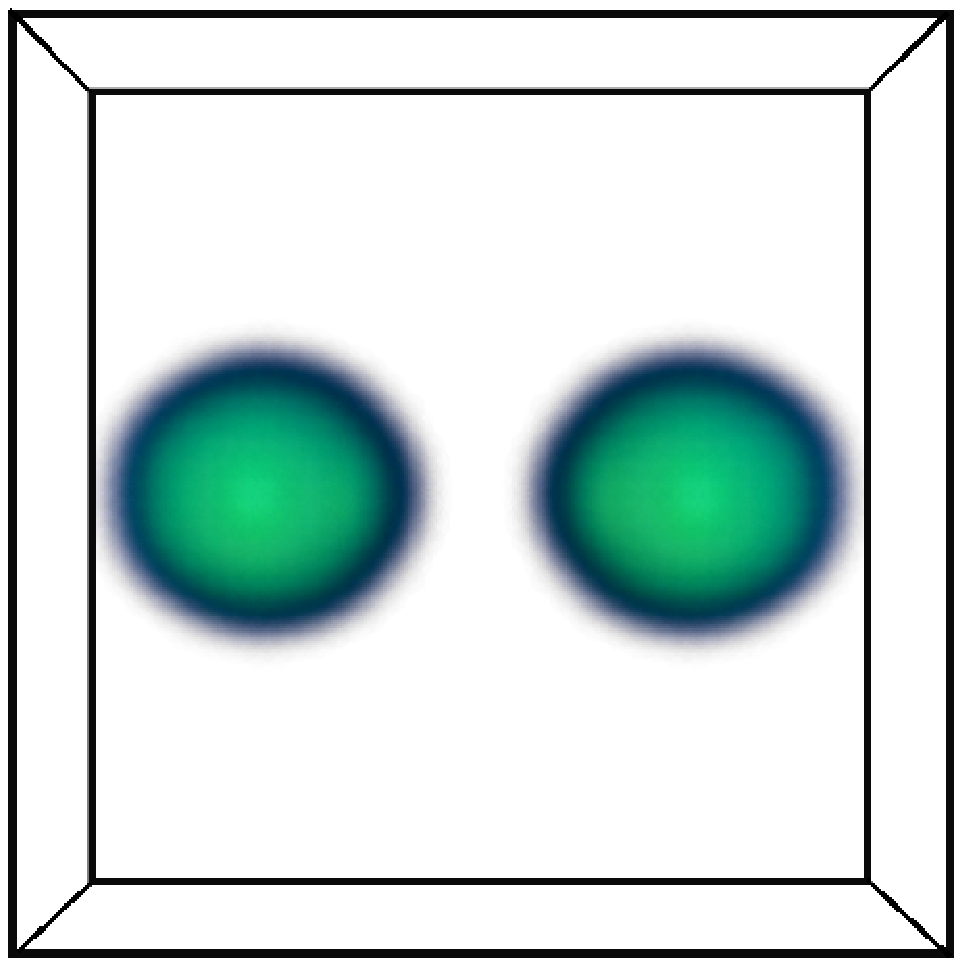} \\
\caption{\label{fig6a} (Colour online) Symmetric fission dynamics from the compound nucleus, which corresponds to the heavy-ion reaction shown in Eq.~\eqref{fission2}, is simulated.
A calculation box is fixed to 48$\times$48$\times$24~fm$^3$.
Snapshots at 0.0$\times$10$^{-22}$~s, 8.3$\times$10$^{-22}$~s, 16.6$\times$10$^{-22}$~s and 24.9$\times$10$^{-22}$~s are shown.
The distance $R_0$, which is deduced from the excitation energy, is equal to 11.8~fm.
Note that the orthogonalization is applied to the initial state before starting TDDFT calculations, because the two initial nuclei are slightly overlapped. }
\end{figure}

Concerning the fission dynamics, here we take the case of $b=4$~fm.
For instance we consider the symmetric fission for $^{296}{\rm Lv}$, which corresponds to the pure TDDFT product.
The distance $R_0 =$11.8~fm is deduced from the excitation energy of the compound nucleus ($^{296}{\rm Lv}$), where trial TDDFT calculations with several $R_0$ were performed to identify $R_0$. 
The fission dynamics
\begin{equation} \label{fission2} \begin{array}{ll}
^{248}{\rm Cm}~+~^{48}{\rm Ca} ~\to~ ^{296}{\rm Lv} ~\to~  ^{148}{\rm Ce} ~+~ ^{148}{\rm Ce} 
\end{array} \end{equation} 
is shown in Fig.~\ref{fig6a}.
The magnitude of $R_0$ is related to the difficulty of fission, as well as the total time-scale of fission.
In addition $R_0$ becomes larger for lower incident energies.
Note that $R_0$ can be larger than the touching distance in which case the fission is suggested to be impossible.
In this sense this method is applicable to the fission appearing in heavy-ion collisions.
Although the duration of fission shown in Fig.~\ref{fig6a} is quite short (similar to the typical duration time of low-energy heavy-ion reactions), the additional time is necessary to realize the initial state.
The total duration time is expected to be significantly longer than 10$^{-21}$~s, because the initial state shown in Fig.~\ref{fig6a}, which cannot be realized in the standard TDDFT at the least, cannot be easily realized.

\section{Summary}  \label{sec3}
A procedure of obtaining fission dynamics of excited compound nuclei has been presented, where the reproduction of microscopic fission dynamics was a long standing problem in nuclear theory (among a few preceding works on microscopic fission dynamics, see J. W. Negele~\cite{negele}).
The presented method allows us to have a self-consistent treatment of fission dynamics.
The obtained dynamics treating the excited states under the strict total energy conservation is diabatic, which is essentially different from the adiabatic fission dynamics.

For the proposed method it is remarkable that the initial condition of fission dynamics is uniquely determined without having any intentional settings; i.e., it is automatically determined for a given set of $A_R$, $Z_R$, $^{A_1}Z_1$, $^{A_2}Z_2$ and $E_R$.
As is seen in the comparison between Figs.~\ref{fig5a} and \ref{fig6a}, the presented fission dynamics is hidden if only pure TDDFT calculations are utilized.
In this sense a new point of this method is to choose an ideal configuration (which is rarely realized in most cases if the excitation energy is not so high) as the initial state.
As a matter of cause, a different choice of $^{A_1}Z_1$ and $^{A_2}Z_2$ brings about different fission dynamics even from an identical compound nucleus.  
It is also worth noting that this method with taking into account many different initial conditions can be utilized to distinguish whether fission arising from the collective dynamics can appear or not.
Even though the proposed procedure might not be the ultimate solution for investigating fission dynamics, the obtained dynamics actually extracts important aspects of fission dynamics (for example, see the time evolution of the neck).

It was suggested by the calculation shown in Fig.~\ref{fig6a} that symmetric fission of the compound nucleus ($^{296}{\rm Lv}$) is possible in the collision: $ ^{248}{\rm Cm}~+~^{48}{\rm Ca}$ with an incident energy $E=$268~MeV.
The fission process itself takes only a few 10$^{-21}$~s, but the preformation of the initial state might take significantly longer.
The estimation of the total duration time for fission process is a future problem. \\

This work was supported by the Helmholtz alliance HA216/EMMI. 
The authors thank Prof. J. A. Maruhn for reading this manuscript carefully.

\end{document}